\documentclass[12pt,amsmath,amssymb,aps
notitlepage,
 reprint,
 superscriptaddress
]{revtex4-2}
\usepackage{graphicx}
\usepackage{epsfig}
\usepackage{color} 

\newcommand{\ket}[1]{\vert #1 \rangle}

\usepackage{epstopdf}

\begin{document}

\preprint{APS/123-QED}

\title{Random Apollonian networks with tailored clustering coefficient}

\author{Eduardo M. K. Souza}
\affiliation{
Departamento de F\'isica, Universidade Federal de Sergipe, 49100-000 S\~{a}o Crist\'{o}v\~{a}o, Sergipe, Brazil
}%

\author{Guilherme M. A. Almeida}
\affiliation{
Instituto de F\'isica, Universidade Federal de Alagoas, 57072-900 Macei\' o, Alagoas, Brazil
}%

\date{\today}

\begin{abstract}
We introduce a family of complex networks that interpolates between 
the Apollonian network and its binary version, recently introduced in [Phys. Rev. E \textbf{107}, 024305 (2023)], via random removal of nodes. The dilution process allows the clustering coefficient to vary from $C=0.828$ to $C=0$ while maintaining the behavior of average path length and other relevant quantities as in the deterministic Apollonian network. Robustness against the random deletion of nodes is also reported on spectral quantities such as the ground-state localization degree and its energy gap to the first excited state. The loss of the 
$2\pi / 3$ rotation symmetry as a tree-like network emerges is investigated in the light of the hub wavefunction amplitude. 
Our findings expose the interplay between the small-world property and other distinctive traits exhibited by Apollonian networks, as well as their resilience against random attacks. 
\end{abstract}


\maketitle

\textit{Introduction.} Network theory is the backbone
of modern research on complex systems \cite{bocca}.
The so-called complex networks can be applied in a variety of fields such as 
health sciences \cite{firth}, the internet \cite{dorog}, social networks \cite{new2}, transportation \cite{latora}, quantum systems \cite{alme, biamonte19}, and more. 
A pivotal goal in exploring complex networks is to disclose universal properties that can accurately describe the physics of 
seemingly distinct systems.

Paradigmatic complex networks include the
Watts-Strogatz \cite{watts}, Barab\'asi-Albert \cite{bara1}, Erd\"os-R\'enyi \cite{eren}, and Apollonian networks \cite{andrade}. These structures have been explored for decades, shedding new light on various kinds of cooperative physical phenomena.
Some of these networks share
the small-world property
\cite{watts} implying that any pair of nodes are within reach by a small number of hops. 
Formally, this property is valid when
the average shortest path length $l \propto \log N$, $N$ being the total number of nodes,
and the clustering coefficient is relatively high, compared to that of a regular network. 

Apollonian networks (ANs) \cite{andrade} have a special appeal as in addition to being small-world, are self-similar, Euclidean, and scale-free, which means that the connectivity degree distribution $P(k)$
obeys a power law. Because of 
its ubiquitous topology, the AN offers an
interesting platform to explore, e.g.,  
condensed-matter phenomena \cite{andrade2,cardoso,oliveira,oliveira2} and quantum walks \cite{xu08,alme}.

%

\begin{figure}
\includegraphics[width=0.40\textwidth]{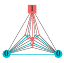}
	\caption{(a) Third generation ($n = 3$) of the AN. The binary ANs \cite{nos} are highlighted by the round and square nodes corresponding to bit 0 and 1, respectively. Here we consider the whole AN and assign a probability $p$ that a node is removed (and so its edges) from the bit-0 subset. Thus, by controlling that parameter we go from the standard AN ($p=0$) to the bit-1 subnetwork ($p=1$).  
 }
	\label{fig1}
\end{figure}

The AN is inspired by the old
circle-packing problem of 
Apollonius.
The network is generated in a simple recursive way. It starts (generation $n=0$) with three nodes forming a triangle.
Subsequent generations are obtained by 
adding a new node inside 
each
triangular face and connecting it to the corresponding nodes occupying the corners.
Figure \ref{fig1} shows the third generation ($n = 3$) of the AN. 

Recently, a subclass of those networks -- coined binary ANs -- was introduced by us in Ref. \cite{nos}. Based on the concept of the binary Pascal's triangle, a bit (0 or 1) can be associated to each node of the AN 
as a result of a modulo-2 addition performed on the bits
linked to the nodes that contain it. 
Applying this scheme from generation $n=0$, one can easily realize that 
a nontrivial output is obtained only when
one bit is different from the other two (as depicted in Fig. \ref{fig1}). 
By removing all the edges connecting nodes having opposite bits, two 
tree-like (and therefore bipartite) networks
take form.
 Interestingly, they obey their own growth rules and inherit many of the characteristics of the original AN such as the scale-free and small-world properties but display zero clustering \cite{nos}.  
 
For this reason, the binary ANs are not strictly classified as small-world networks. \cite{watts}.
To fulfill the condition, they would need to have a clustering coefficient $C$ at least higher than that of a random network, as observed in many real-world networks. 
This coefficient can be decisive for, e.g., influencing
the spread velocity of a disease outbreak \cite{zhou05} or rendering a network more vulnerable to attacks 
\cite{holmes02}.
The celebrated Watts-Strogatz model \cite{watts}, for example, interpolates between a regular ring lattice (high clustering) and a random graph (short path length). 
%
Indeed, 
metrics employed to assess 
the small-world property are often based on the interplay between clustering and path length \cite{teleford11, boguna20}.
%
%

In this letter we define
a network that interpolates between the standard AN and one of its binary subnetworks by means of a procedure involving random removal of nodes. This allows us to 
vary the clustering coefficient from $C=0.828$ \cite{andrade} to $C=0$ \cite{nos}, respectively, while maintaining the scaling of the path length. 
Rather than quantitatively evaluate
the small-worldness of the network, we want here
to address how the variation of $C$ affects other physical properties of interest. 
On the one hand, this approach allows us to 
weight down 
to what degree the small-world property shapes the overall behavior of the network. On the other, it unveils
how the unique set of characteristics of the AN emerge from its constituent parts, as well as their resilience against random attacks.  

%


\textit{Diluted Apollonian networks}. 
Let us build a variation of the AN by defining a probability $p$ that
a node belonging to the bit-0 subnetwork (highlighted as round nodes in Fig. \ref{fig1}) is removed alongside all the edges attached to it. All the nodes independently respond to same removal probability. Hence, a \textit{diluted} version of the AN is obtained after removal of $pN_n^{(0)}$ nodes of the bit-0 subnetwork on average. 
Given the number of nodes of the AN, 
0-bit, and 1-bit subnetworks are, respectively, 
$N_{n}^{(\mathrm{AN})} = (3^{n}+5)/2$,
$N_{n}^{(0)} =(3^{n}+(-1)^{n}+6)/4 $, and $N_{n}^{(1)} = (3^{n}-(-1)^{n}+4)/4$ \cite{andrade, nos}, the diluted AN will have 
$N_{n} = N_{n}^{(\mathrm{AN})} - p N_{n}^{(0)}$ nodes.
We mention that setting the $p=0$ limit 
to the bit-0 subnetwork instead (by diluting the bit-1 nodes) provides with qualitatively similar results. Hence, it will not be addressed here.




\textit{Results.} Our analysis is 
based on the network adjacency matrix. 
We consider undirected edges and define a
symmetric adjacency matrix $\textbf{A}$ that contains all the information concerning the network topology.
Its elements $A_{ij}=1$ when there is an edge connecting nodes $i$ and $j$, and $A_{ij}=0$ otherwise. Note that a general tight-binding Hamiltonian can be defined in terms of the adjacency matrix with a proper energy unit. 
We will first discuss structural properties evaluated on $\textbf{A}$ in the local basis, followed by its spectral properties. 
%
Calculations are done for generations $n=5,6,7,8$ and averaged over $500,200,100,50$ independent samples, respectively. 

\begin{figure}
	\includegraphics[width=0.49\textwidth]{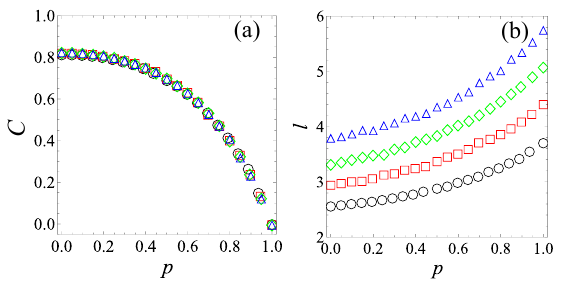}
	\caption{(a) Clustering coefficient $C$ and (b) average shortest path length $l$ versus node removal probability $p$. Results are shown for generations $n=5$ (circles), $n=6$ (squares), $n=7$ (diamonds), and $n=8$ (triangles). In the AN limit ($p=0$) $C=0.828$ \cite{andrade} whereas $C=0$ when $p=1$, which reflects the tree-like structure of the binary ANs \cite{nos}. We note that $l \propto (\log N_{n})^{3/4}$ for all $p$ and $n$.  
 }
	\label{fig2}
\end{figure}

Let us begin with the clustering coefficient $C$, which
will guide us throughout this letter.  
The \textit{local} clustering coefficient is defined as
\begin{equation}
C_{i}= \frac{1}{k_{i}(1-k_i)} \sum_{jl} A_{ij} A_{jl} A_{li},
\end{equation}
with the degree $k_{i}=\sum_{j} A_{ij}$ being the number of edges attached to node $i$ and $k_{i} \ne 0,1$ ($C_{i}=0$ otherwise). 
The clustering coefficient $C\in [0,1]$ is the average $C=\frac{1}{N_{n}} \sum_{i} C_{i}$ \cite{watts} and thus is proportional to the number of triangles in the network. 
%
%
Figure \ref{fig2}(a) shows $C$ as a function of $p$.
It goes from $C=0.828$,
which corresponds to the AN ($p=0$) \cite{andrade}, 
to $C=0$ when the bit-1 subnetwork ($p=1$) is achieved \cite{nos}. 
The likelihood of broken links between both binary ANs increases with $p$, diminishing the density of triangles in the network. 
%
However, we observe that $C$ remains considerably high, close to the level that corresponds to the standard AN,
for probabilities as high as $p\approx 0.5$.
Also note that the system size $N_n$ does not affect the curve. 
%

The other
parameter that accounts for the small-world effect is the average shortest path length 
defined as
\begin{equation} \label{eq:lm}
l = \frac{1}{N_n(N_n-1)}\sum_{i \neq j} d_{i,j},
\end{equation}
where $d_{i,j}$ is the length of the shortest path between nodes $i$ and $j$.
As can be seen in Fig. \ref{fig2}(b), $l$ slightly increases with $p$, an expected tendency given the density of triangles is diminishing.   
The important feature to address here is how $l$ scales with $N_{n}$.  Assuming $l \propto (\log N_{n})^{\beta}$ we find the exponent $\beta=3/4$ for all values of $p$.
Thus, the whole family of diluted ANs falls in an intermediate class between small ($l \propto \log N_{n}$) and ultrasmall [$l \propto \log(\log N_{n})$] networks.

\textcolor{black}{So far we realize that the small-world character of the AN is resilient to the random deletion of nodes. 
In addition, only the clustering coefficient $C$ is affected by the dilution procedure. This suggests that the binary subnetworks are key building blocks of the standard AN. 
} 
%

%
%

%

\begin{figure}
	\includegraphics[width=0.49\textwidth]{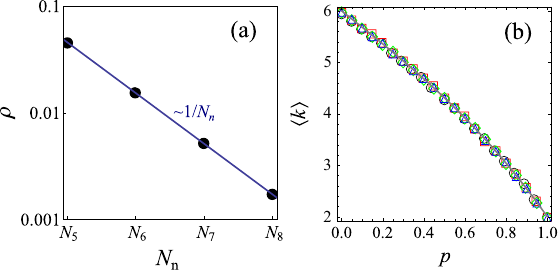}
	\caption{(a) Connectance $\rho$ versus network size $N_n$ in log-log scale for the representative value $p=0.5$, where
 $N_5 = 124 - 62p$, $N_6=367 - 184p$, $N_7=1096 - 548p$, and $N_8=3283 - 1642p$. The symbols
represent the numeric data and the solid line depicts a function $\rho \propto N_n^{-1}$. This scaling behavior renders the convergence of the average degree $\langle k \rangle$, plotted in (b) against $p$ for generations
$n=5$ (circles), $n=6$ (squares), $n=7$ (diamonds), and $n=8$ (triangles). The gray line is the function $\langle k \rangle(p) = 6- 2\gamma p + 2(\gamma-2)p^{3}$. Note that $\langle k \rangle(0)=6$ (AN) and $\langle k \rangle(1)=2$ (1-bit subnetwork), as expected.}
	\label{fig3}
\end{figure}

Small-world networks with a power-law degree distribution are generally robust against attacks of that nature due to the presence of hubs \cite{cohen1}. 
In fact, the AN and the binary ANs are scale-free, a trait also portrayed by the diluted AN.
The cumulative degree distribution is defined as
\begin{equation} \label{eq:eqpk}
P(k) = \sum_{k' \ge k} \frac{m(k',n)}{N_n},
\end{equation}
where $m(k,n)$ is the number of sites with a certain degree $k$ at generation $n$. 
Here we indeed observe that the diluted
AN is scale free, as characterized by 
the asymptotic form $P(k) \propto k^{-\gamma}$, with $\gamma = \ln 3/\ln 2 = 1.585$ for all $p$.
%
%

The scale-free attribute implies in the existence of a larger number 
of low-degree nodes that form dense communities
linked through local hubs. In the standard AN most of the nodes
have degree $k=3$, precisely $3^{n-1}$ nodes 
for a given generation $n>1$. In the bit-1 subnetwork
$[3^{n-1}+(-1)^{n-1}]/2$ ($n>2$) nodes have degree $k=1$ 
which signalizes its dendritic pattern. 
Both limiting networks are sparse
in the sense that their average degree $\langle k \rangle (p)$ converges to a constant, namely 
$\langle k \rangle (0) \rightarrow 6$ (AN) and $\langle k \rangle (1) \rightarrow 2$ (1-bit AN) as $N_{n} \rightarrow \infty$ \cite{nos}.

The convergence of $\langle k \rangle$ can be derived from the relationship 
$\langle k \rangle = (N_{n}-1)\rho$, where $\rho=B/B_{\mathrm{max}}$ is the connectance, that is the ratio
between total number of edges $B=\sum_{i,j}A_{ij}/2$ and its maximum possible value $B_{\mathrm{max}}=N(N-1)/2$ (which corresponds to a fully connected network). 
For the whole range of $p$ we obtain $\rho \propto N_{n}^{-1}$, as depicted in Fig. \ref{fig3}(a). 
Hence, the diluted AN is 
also sparse, with $\langle k \rangle$ 
varying with $p$ as seen in Fig. \ref{fig3}(b).
Such behavior can be generated by
the polynomial $\langle k \rangle (p) = 6 - 2\gamma p + 2(\gamma-2)p^{3} + O(N_{n}^{-1})$. 
Interestingly, so far as the structural properties discussed above are concerned, the diluted AN
behaves much like the standard AN (up to intermediate values of $p$) despite having a distinct topology. For instance, when $p=0.5$ we have $\langle k \rangle = 4.32$ and the small-world property still preserved with $C\approx 0.7$ and $l\propto (\log N_n)^{3/4}$.


\begin{figure}
	\includegraphics[width=0.49\textwidth]{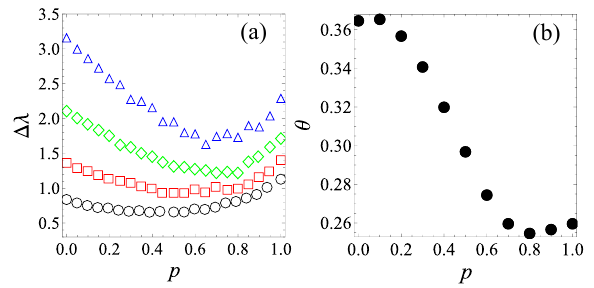}
	\caption{(a) Energy gap $\Delta \lambda$ versus $p$ evaluated at generations
$n=5$ (circles), $n=6$ (squares), $n=7$ (diamonds), and $n=8$ (triangles). (b) Exponent of the power-law scaling $\Delta \lambda \propto N_n^{\theta(p)}$.}
	\label{fig4}
\end{figure}

Let us now move on to discuss some spectral properties of the network, starting with the energy (eigenvalue) gap between the ground and the first excited states, $\Delta \lambda$. 
This is an important parameter that can be used to identify, e.g., quantum phase transitions \cite{presilla19}. It also plays a pivotal role in the critical 
behavior of Bose-Einstein condensates \cite{oliveira2}. 
The energy gap versus $p$ is presented in Fig. \ref{fig4}(a). It is found to obey the power-law scaling $\Delta \lambda \propto N_n^{\theta(p)}$, with the exponent $\theta(p)$
behaving as shown in Fig. \ref{fig4}(b).
The limiting cases $\theta(0)=0.34$ (AN) $\theta(1) = 0.26$ (1-bit subnetwork) can be found elsewhere \cite{oliveira2, nos}. 
The scaling behavior of the gap does not undergo
relevant changes and its divergence relates
to the ever increasing degree of the nodes belonging 
to the inner generations \cite{las,goh}. It also deserves notice that the 1-bit subnetwork retains the original AN hub (see Fig. \ref{fig1}). In the 0-bit subnetwork, contrarily, 
$\theta = 0.06$ as found in \cite{nos}.

\begin{figure}
	\includegraphics[width=0.4\textwidth]{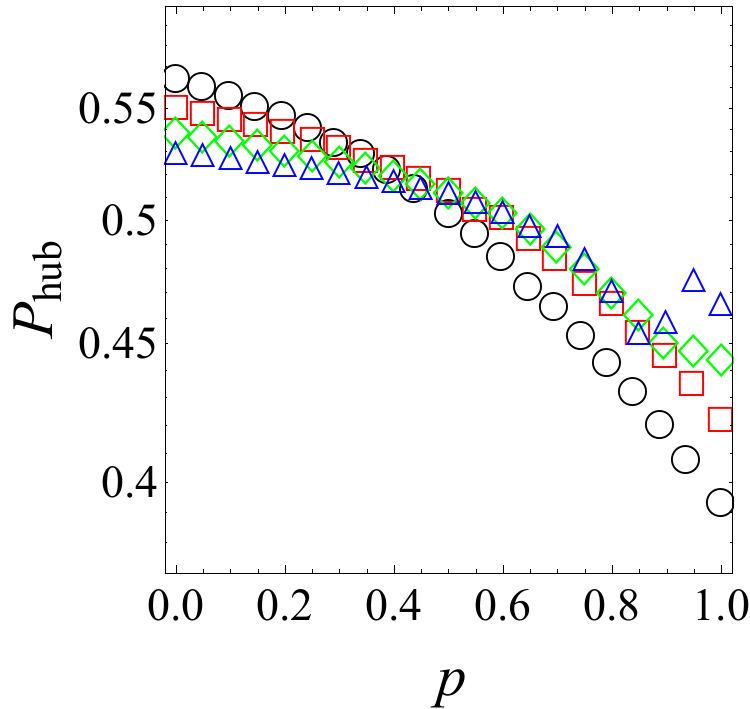}
	\caption{Probability amplitude $P_{\mathrm{hub}}=|v_{\mathrm{hub}}|^2$ in the ground state versus $p$ for various $N_n$, where
$n=5$ (circles), $n=6$ (squares), $n=7$ (diamonds), and $n=8$ (triangles).}
	\label{fig5}
\end{figure}

\begin{figure}
	\includegraphics[width=0.49\textwidth]{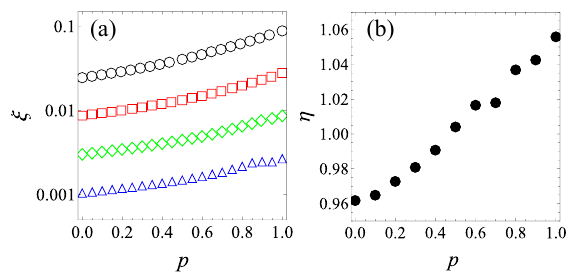}
	\caption{(a) Ground-state participation ratio $\xi$ versus $p$ for generations
$n=5$ (circles), $n=6$ (squares), $n=7$ (diamonds), and $n=8$ (triangles). (b) Exponent $\eta$ corresponding to the scaling $\xi \propto N_n^{-\eta}$.}
	\label{fig6}
\end{figure}

Last, we 
turn our attention to the 
ground-state eigenvector itself $\ket{v}=\lbrace v_{1},v_{2},v_{3},\ldots,v_{N_n} \rbrace$, which happens to feature remarkable localization properties. In scale-free networks with undirected edges the extreme eigenvectors are strongly localized at the hub \cite{goh}. 


In Fig. \ref{fig5} we plot the occupation probability at the hub versus $p$ with respect to $\ket{v}$, that is $P_{\mathrm{hub}}=|v_{\mathrm{hub}}|^{2}$. 
%
%
We identify a transition between two regimes at about $p= 0.4$. Below this value,
$P_{\mathrm{hub}}$ decreases as $N_{n}$ grows.
This happens because even though $\ket{v}$ is always strongly localized at the hub, reconfiguration of the network upon addition of a large number nodes at every new generation follows a \textit{self-similar} pattern in the AN ($p= 0$). The unfolding $2\pi / 3$ rotation symmetry produce a highly structured and degenerate
energy spectrum featuring a mix between extended and localized states, the degree of which typically depends on the node degree $k_i$ \cite{cardoso,alme}. 
Furthermore, for all $n$ considered the amplitude decays to a certain point with increasing $p$.  
This trend can be associated to the rotation symmetry break and decrease in $k_{\mathrm{hub}}$.


The 1-bit subnetwork, one the other hand,
is bipartite \cite{nos}. 
Nodes can be grouped into two subsets, $S_1$ and $S_2$, with no edges connecting nodes within the same subset. 
This arrangement satisfies $\sum_{i\in S_1}|v_{i}|^2=\sum_{i\in S_2}|v_{i}|^2=1/2$ \cite{souzabi}. 
As $N_n$ increases, this constraint
and the scale-free property of the network render extreme eigenstates carrying approximately 50$\%$ of their amplitude
at the hub as seen in Fig. \ref{fig5} (see also Ref. \cite{nos}). The remaining is mostly distributed among its first low-degree neighbors.
%
This tendency is further explored in Appendix \ref{ap}.
A similar argument can be used to explain the apparent convergence $P_{\mathrm{hub}}\rightarrow 1/2$ in the AN case ($p=0$) as $N_n$ increases. It can be verified numerically that the components $|v_i|$ approach zero given $A_{\mathrm{hub},i}=0$ in the adjacency matrix. Despite the lack of bipartite symmetry in the AN, its $2\pi / 3$ rotation symmetry and community structure ruled by sets of local hubs \cite{alme,cardoso} contribute to the onset of a gapped ground state (cf. Fig. \ref{fig4}) resembling that of a very dense star graph (see Fig. \ref{fig7} and discussion in Appendix \ref{ap}) for higher generations $n$. 

Next, we want to analyze the localization degree of $\ket{v}$, whic can be done through the inverse participation ratio defined as
$\xi = \frac{1}{N_{n}\sum_{i} |v_{i}|^{4}}$. It yields $\xi=N_n^{-1}$ for fully localized states and $\xi=1$ for extended states. Assuming 
$\xi \propto (N_{n})^{-\eta(p)}$, the decay coefficient $\eta(p)$ quantifies the localization degree. In  
Figs. \ref{fig6}(a) and
\ref{fig6}(b) we show both 
$\xi$ and $\eta$ against $p$, respectively, the latter being $\eta \approx 1$
throughout. 

We confirm that in the midst of the departure from the AN rotation symmetry toward the tree-like form as the clustering coefficient fades, the ground state remains strongly localized. And the state has its highest amplitude at the node featuring the highest degree, as expected. 

\textit{Conclusions.}
We have seen that the AN is quite robust against the random removal of nodes of the binary partition. Network properties such as the shortest path length 
$l \propto (\log N_{n})^{\beta}$ and the degree distribution 
$P(k) \propto k^{-\gamma}$ maintained their 
exponents $\beta=3/4$ and $\gamma=\ln 3/ \ln 2$ for all $p$ despite $C\rightarrow 0$.
We also observed no significant changes to the exponent $\theta(p)$ associated to the gap between the ground and first excited states 
$\Delta \lambda \propto N_n^{\theta(p)}$.
One contributing factor to this is that the 1-bit subnetwork
includes the hub (central node) of the original AN (see Fig. \ref{fig1}).
The scaling of the gap influences, for instance, the critical temperature of a Bose-Einstein condensation in the AN \cite{oliveira2}.  


%
In terms of the small-world property that presuppose a short $l$ and high $C$, we remark
that $C$ remains close to the level corresponding to the AN, $C=0.828$, for $p$ values as high as $p=0.5$. 
This is interesting if we contemplate how 
the AN topology is modified under the random attacks to the 0-bit partition. 
The diluted AN is always sparse with the average degree obeying $\langle k \rangle (p) = 6 - 2\gamma p + 2(\gamma-2)p^{3} + O(N_{n}^{-1})$, where $\gamma=\ln 3/\ln 2$.
At $p=0.5$, $\langle k \rangle = 4.32$, indicating a dramatic change in the network topology. 

We highlight that the symmetry underlying the diluted AN stands halfway between the $2\pi / 3$ rotation symmetry portrayed by the AN and the dendritic profile of the 1-bit partition. This pattern  
has been shown to be crucial in determining the hub amplitude in the ground state. Yet, its overall localization degree as computed via the inverse participation ratio yields 
$\xi \propto (N_{n})^{-\eta(p)}$, with $\eta(p)\approx 1$ for all $p$. This, once again, confers robustness to the AN. 
   
The binary AN subnetworks \cite{nos} are indeed key partitions that carry the essence of the standard AN \cite{andrade}. Interpolating between 
them broadens the range of applicability, specially if a tunable clustering coefficient $C$ is desired. Further studies directed toward the modeling of real-world networks and other complex systems using the diluted AN class are timely.  
 
The authors thank A. M. C. Souza and L. K. Souza for sharing key insights. 
This work was supported by CNPq (Brazilian agency) and FAPEAL (Alagoas State agency).

\appendix
\section{Competing hubs in dendritic networks \label{ap}} 

In this section we devise a toy model
to explain why the probability amplitude of the 1-bit subnetwork hub $P_{\mathrm{hub}}=|v_{\mathrm{hub}}|^2$
in the ground state tends to $1/2$ 
as $N_n$ increases.

\begin{figure}
	\includegraphics[width=0.40\textwidth]{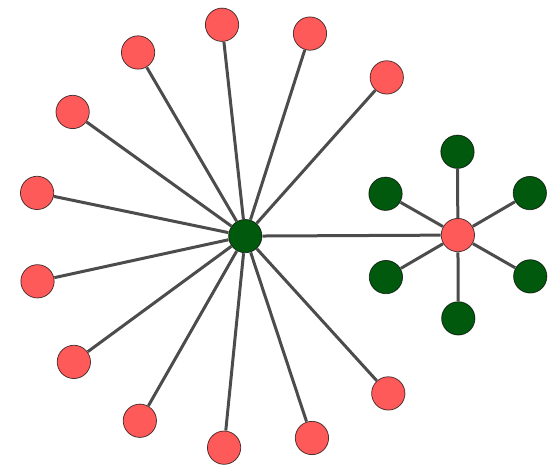}
	\caption{Two star graphs coupled via their central nodes $\ket{c_1}$ and $\ket{c_2}$. Here, $k_1 = 12$ and $k_2=6$. 
Bright and dark nodes indicate the disjoint subsets that are formed due to the bipartite symmetry of the network.}
	\label{fig7}
\end{figure}

Let us consider two star networks coupled via their respective central nodes (local hubs) $\ket{c_1}$ and $\ket{c_2}$, with degrees $k_{1}$ and $k_2$, as depicted in Fig. \ref{fig7}. The nodes are marked with different colors to highlight the disjoint sets $S_1$ and $S_2$ due to the bipartite property of the network. 
The
total number of nodes is thus $k_1+k_2+2$. Once again, we assume a symmetric adjacency matrix $\mathbf{A}$ with non-zero elements $A_{ij}=1$ only if an edge connects nodes $i$ and $j$.

Each star network is itself bipartite as well. By carrying out a partial diagonalization, each one delivers $k_m-1$ ($m=1,2$) zero-energy eigenstates $\ket{\phi_{m,i}}$ having \textit{null} amplitude at the star center $\ket{c_m}$ and two others with energies $E=\pm \sqrt{k_m}$:
\begin{equation}
\ket{\psi_m^{\pm}}=\frac{1}{\sqrt{2}}\left( \ket{c_m} \pm \ket{\alpha_m} \right),
\end{equation}
where $\ket{\alpha_m} = \sum_{i=1}^{k_m}k_m^{-1/2}\ket{i_m}$ is an even linear combination of the $k_m$ nodes $\lbrace 
\ket{i_m} \rbrace$ attached to the corresponding star center. 

The components of the non-zero eigenstates are divided such that there is a $50\%$ chance to find a particle at the center or distributed over the other disjoint set, in accordance to \cite{souzabi}. Another theorem \cite{inui94} has it that in every bipartite network there should be at least a number of zero-energy eigenstates that equals the difference between the number of nodes belonging to each disjoint set. And 
these states will have no amplitude in the minority set. 

Returning to the coupled scenario (as in Fig. \ref{fig7}), we realize that the zero and non-zero eigenstates as obtained above do not mix because both parts are connected via their local hubs. That is, $\langle \phi_{m,i} \vert \textbf{A} \vert \psi_{m'}^{\pm} \rangle = 0$. As such, the four states $\ket{\psi_{m}^{\pm}}$ span their own eigenspace. We now proceed with a change of basis to the set $\lbrace \ket{\alpha_1}, \ket{c_1},\ket{c_2}, \ket{\alpha_2} \rbrace$ to express 
the effective adjacency matrix (written in that order) as
\begin{equation}
\mathbf{A_{\mathrm{eff}}}=
\begin{pmatrix}
0 & \sqrt{k_1} & 0 & 0 \\
\sqrt{k_1} & 0 & 1 & 0 \\
0 & 1 & 0 & \sqrt{k_2} \\
0 & 0 & \sqrt{k_2} & 0 
\end{pmatrix}.
\end{equation}

Surprisingly, we are able to express the coupled star network system of Fig. \ref{fig7} as a linear chain in the state sector that contains each star center. Similar structures can be found within networks that exhibits
dendritic (tree-like) patterns such as the binary ANs \cite{nos}. Solving the eigenvalue equation for $\mathbf{A_{\mathrm{eff}}}$, we obtain
\begin{equation} \label{energy}
E_{\mu,\nu} = \frac{\mu}{\sqrt{2}}\sqrt{1+k_1+k_2+\nu \Omega},
\end{equation}
where $\mu,\nu = \pm 1$ and $\Omega = [(k_1-k_2+1)^2+4k_2]^{1/2}$.
The corresponding eigenvectors have the form
\begin{equation} \label{vector}
\ket{E_{\mu,\nu}}=\mathcal{N}\begin{pmatrix}
1 \\
\frac{E_{\mu,\nu}}{\sqrt{k_1}} \\
\frac{(E_{\mu,\nu}^2-k_1)}{\sqrt{k_1}} \\
\frac{\sqrt{k_2}(E_{\mu,\nu}^2-k_1)}{E_{\mu,\nu}\sqrt{k_1}} 
\end{pmatrix},
\end{equation}
where $\mathcal{N}$ is a normalization factor. 

In the effective 4-node network, $\ket{c_1}$ ($\ket{c_2}$) pair up with $\ket{\alpha_2}$ ($\ket{\alpha_1}$) with respect to disjoint bipartite subsets. Accordingly, we must have $|\langle c_m | E_{\mu,\nu} \rangle|^2+|\langle \alpha_{m'} | E_{\mu,\nu} \rangle|^2=1/2$, with $m\neq m'$ \cite{souzabi}. This can be regarded as a competition between one star center and the nodes adjacent to the other in order to fulfill the $50\%$ occupation probability allocated for them.

In a situation in which, say $k_1 \gg k_2$, the eigenvalues [Eq. (\ref{energy})] approach $E_{\mu,+}\approx \mu \sqrt{k_1}$ and $E_{\mu,-}\approx \mu \sqrt{k_2}$. 
After some algebraic manipulation on 
Eq. (\ref{vector}) we find that
the extreme eigenstates 
$\ket{E_{\mu,+}}\approx \ket{\psi_{1}^{\mu}}$ (they are indeed associated to the node with the highest degree). Therefore, 
 $P_{\mathrm{hub}}\equiv|\langle c_1 \vert E_{\mu,+} \rangle|^2 \rightarrow 1/2^{-}$ as $k_1 \rightarrow \infty$.

To summarize, whenever one of the star networks becomes very dense, it effectively splits apart from the other. 
Notwithstanding that the 1-bit AN subnetwork has a much more complex topology, 
the above picture captures the essence of the behavior seen in Fig. \ref{fig5} in the unclustered limit ($p=1$) as $N_n$ grows. 


\end{document}